\documentclass[solaromanenum]{solarphysics}
\usepackage[hyperref,optionalrh,solaromanenum]{spr-sola-addons} 
\usepackage{graphicx}                    
\usepackage{color}                       
\usepackage{breakurl}                         
\usepackage[T1]{fontenc}

\usepackage{multirow}

\begin{document}
\begin{article}
\begin{opening}

\title{Quasi-Periodic Pulsations Detected in Ly\,$\alpha$ and Nonthermal Emissions During Solar Flares}


\author[addressref={aff1,aff2},email={leilu@pmo.ac.cn}]{\inits{L.}\fnm{Lei}~\lnm{Lu}\orcid{0000-0002-3032-6066}}
\author[addressref={aff1,aff2},corref,email={lidong@pmo.ac.cn}]{\inits{D.}\fnm{Dong}~\lnm{Li}\orcid{0000-0002-4538-9350}}
\author[addressref={aff1,aff3},email={}]{\inits{Z.J.}\fnm{Zongjun}~\lnm{Ning}\orcid{0000-0002-9893-4711}}
\author[addressref={aff1,aff3},email={lfeng@pmo.ac.cn}]{\inits{L.}\fnm{Li}~\lnm{Feng}\orcid{0000-0003-4655-6939}}
\author[addressref={aff1,aff3},email={}]{\inits{W.Q.}\fnm{Weiqun}~\lnm{Gan}}

\runningauthor{L. Lu et al.}
\runningtitle{QPPs Detected in Ly\,$\alpha$ and Nonthermal Emissions During Solar Flares}

\address[id=aff1]{Key Laboratory of Dark Matter and Space Astronomy, Purple Mountain Observatory, CAS, Nanjing 210033, PR China}
\address[id=aff2]{CAS Key Laboratory of Solar Activity, National Astronomical Observatories, Beijing 100101, China}
\address[id=aff3]{School of Astronomy and Space Science, University of Science and Technology of China, Hefei 230026, PR China}
\begin{abstract}
We report quasi-periodic pulsations (QPPs) with double periods
during three solar flares (viz. SOL2011-Feb-15T01:44,
SOL2011-Sep-25T04:31, SOL2012-May-17T01:25). The flare QPPs were
observed from light curves in Ly\,$\alpha$, hard X-ray (HXR) and
microwave emissions, with the Ly\,$\alpha$ emission recorded by the
\emph{Geostationary Operational Environmental Satellite}, the HXR
emission recorded by the \emph{Reuven Ramaty High-Energy Solar
Spectroscopic Imager} and the \emph{Fermi Gamma-ray Burst Monitor},
and the microwave emission recorded by the \emph{Nobeyama Radio
Polarimeters} and \emph{Radioheliograph}. By using the Markov chain
Monte Carlo (MCMC) method, QPPs with double periods of about two
minutes and one minute were first found in the Ly\,$\alpha$
emission. Then using the same method, a QPP with nearly the same
period of about two minutes was also found in HXR and microwave
emissions. Considering the possible common origin (nonthermal
electrons) between Ly\,$\alpha$ and HXR/microwave emission, we
suggest that the two-minute QPP results from the periodic
acceleration of nonthermal electrons during magnetic reconnections.
The ratio between the double periods in the Ly\,$\alpha$ emission
was found to be close to two, which is consistent with the
theoretical expectation between the fundamental and harmonic modes.
However, we cannot rule out other possible driving mechanisms for
the one-minute QPPs in HXR/microwave emissions due to their
relatively large deviations.
\end{abstract}


\end{opening}
\section{Introduction}
\label{S-Introduction}
A solar flare  is a sudden localized
brightening in the solar atmosphere, corresponding to a rapid and
violent energy release through the widely accepted
magnetic-reconnection process \citep{Masuda94,Benz17,Yan18,Lid21}.
The flare emission often exhibits quasi-periodic temporal
variations, which are referred to as quasi-periodic pulsations
(QPPs) in solar flares \citep[see][for reviews]{Nakariakov09,Van16}.
Based on the time-series analysis, QPPs have been observed from
light curves in various wavebands, from radio waves
\citep{Kolotkov15,Nakariakov18,Yu19,Karlicky20} through ultraviolet
(UV) and extreme ultraviolet (EUV) wavelengths
\citep{Brosius16,Li16,Kumar17,Shen18,Dominique18,Miao20} to
soft/hard X-rays (SXR/HXR) channels
\citep{Li08,Ning14,Dennis17,Inglis17,Lid17,Kolotkov18}, and even to
$\gamma$-rays \citep{Nakariakov10,Li20a}. Thanks to spectroscopic
observations with high temporal resolution, the flare-related QPPs
were also found in spectral-line profiles of emission lines, such as
the periodic variations in the line intensity and width,
oscillations of the derived Doppler shift
\citep{Brosius15,Lit15,Li15,Brosius16,Zhang16,Li18}, and the
\emph{Solar Ultraviolet Measurements of Emitted Radiation} (SUMER)
oscillations with strong damping \citep{Ofman02,Wang02,Wang11}.
Understanding of the formation mechanism of QPPs is important for us
to understand the energy release, particle acceleration, and plasma
heating during solar flares, and it may also be useful for the
forecast of solar flares \citep[e.g.][]{Van16,McLaughlin18}.

The characteristic periods of the flare-related QPPs were found to
vary from milliseconds to dozens of minutes
\citep[e.g.][]{Kupriyanova10,Tan10,Dolla12,Shen13,Cho16,Duckenfield18,
Kobanov19,Karlicky20,Hayes20,Li20b}, depending on the different
observation wavebands
\citep{Tan07,Huang14,Dennis17,Nistico17,Li17,Pugh19}. This implies
that the various QPPs are likely to originate from different
physical mechanisms \citep{Nakariakov09,Tan10}, which are still an
open issue \citep[see][for reviews]{McLaughlin18}. Some studies
suggest that the QPPs could be directly interpreted in terms of
magnetohydrodynamic (MHD) waves, such as slow waves, kink waves, and
sausage waves \citep{Anfinogentov15,Wang15,Mandal16,Tian16,Yuan16,
Nakariakov16,Nakariakov19,Nakariakov20}, while some others tend to
associate the QPPs with periodic magnetic reconnections, which could
be either spontaneous or triggered by MHD oscillations
\citep{Kliem00,Chen06,Li15,Guidoni16,Thurgood17,Yuan19}.

The Lyman-$\alpha$ (Ly\,$\alpha$) line at 1216\,{\AA} is a spectral
line of hydrogen, which is formed in the chromosphere
\citep{Canfield81,Allred05}. Since the solar atmosphere contains a
large abundance of hydrogen, the Ly\,$\alpha$ line is thought to be
the strongest emission line in the solar UV spectrum. It was found
that Ly\,$\alpha$ line emission not only shows a significant
intensity enhancement during the solar flare
\citep{Curdt01,Woods04,Milligan12,Milligan14,Milligan16,Hong19,Lu21},
but also shows a clear response to coronal loops \citep{Ishikawa17}
and filament eruptions \citep{Susino18}. Moreover, the increased
Ly\,$\alpha$ emission was found to be closely correlated with
induced currents in the Earth's ionosphere, particularly during
solar high-activity periods \citep[see][]{Milligan20}. Thus study of
the variation of Ly\,$\alpha$ emission is helpful for understanding
the dynamics of the terrestrial environment.

Flare-related QPPs in Ly\,$\alpha$ emission have been reported by
several authors. \cite{Ishikawa17} reported short-period variations
($<$30\,second) in the Ly\,$\alpha$ emission of coronal loops, which
is supposed to be driven by nanoflares. $\approx$3-minute and
$\approx$4.4-minute QPPs were discovered in the full-Sun integrated
Ly-$\alpha$ emission during the impulsive phase of two X-class
flares, both of which were interpreted in terms of flare-induced
acoustic waves \citep[see][]{Milligan17,Milligan20,Li20d}.
$\approx$1-minute QPP is detected in the full-Sun Ly\,$\alpha$
emission during two solar flares, which is attributed to the
repetitive magnetic reconnection \citep{Li20c}. QPPs with both short
($\approx$8.5\,second) and long ($\approx$63\,second) periods were
discovered in the Ly\,$\alpha$ channel during a solar flare, which
are considered as standing fast and slow sausage modes, respectively
\citep{Van11}.

The Ly\,$\alpha$ emission enhancement during solar flares was found
to be highly synchronous with emission enhancement in hard X-ray
\citep{Nusinov06}. The source of Ly\,$\alpha$ emission was further
found to be co-spatial with HXR sources located at flare footpoints
\citep{Rubio09}. Recently, a statistical study made by \cite{Jing20}
shows that the  Ly\,$\alpha$ emission enhancement that appears in
the impulsive phase of a solar flare generally follows the Neupert
effect \citep{Neupert68}. These observations show a close
relationship between Ly\,$\alpha$ and hard X-ray emissions.
Considering the common nonthermal origin between HXR and microwave
emissions \citep{Dulk85,Holt69,Aschwanden87}, a close relation
between the Ly\,$\alpha$ and microwave emission is expected. In the
present study, we investigated QPPs that are observed in
Ly\,$\alpha$, HXR, and microwave emissions during three powerful
flares. This article is organized as follows: Section~2 describes
the observations and method used in this study, Section~3 presents
our primarily results, and Section~4 summarizes the conclusions and
discussion.

\section{Observations and Method}
Three solar flares that occurred on 15 February 2011, 25 September
2011, and 17 May 2012, are selected to investigate the QPPs with
double periods, as shown in Table \ref{tab1}. These flares were
simultaneously recorded by multiple instruments. The \emph{X-Ray
Sensor} \citep[XRS:][]{Hanser96} onboard the \emph{Geostationary
Operational Environmental Satellite} (GOES) observes the full-Sun
integrated soft X-ray emissions in both 0.5\,--\,4\,{\AA} and
1\,--\,8\,{\AA} wavebands. The flux measurements in 1\,--\,8\,{\AA}
have become a standard for classifying solar flares (A, B, C, M, and
X), according to which the three flares under study (one X-class
flare and two M-class flares) can be considered as powerful solar
eruptions.

Starting from 2006, in addition to the XRS, GOES-13 and subsequent
GOES-14, GOES-15, etc., began to carry an \emph{EUV Sensor}
\citep[EUVS;][]{Viereck07}. GOES/EUVS measures the full-Sun EUV
radiations in five wavelength channels, viz., A, B, C, D, and E,
with the E channel targeting on the Ly\,$\alpha$ emission, which
enables us to make a detailed analysis of flare oscillations in the
Ly\,$\alpha$ waveband. The temporal cadence of GOES/EUVS-E
measurements is nominally 10.24\,seconds, but it changes to
12.29\,seconds every third measurement \citep[see][]{Viereck07}.
Given that the QPP analysis requires an uniform time cadence, the
Ly\,$\alpha$ emission measurements were temporally interpolated to
10.24-second intervals.

\begin{table}
\caption{Characteristics of three solar flares studied in this article.}
\label{tab1}
\setlength{\tabcolsep}{2pt}
\begin{tabular}{c c c c c}
\hline
 No. & Flare & GOES class &  NOAA Number & Position  \\
  1 & SOL2011-Feb-15T01:44 & X2.2 & 11158 & [205,-222]  \\
  2 & SOL2011-Sep-25T04:31 & M7.4 & 11302 & [-688,108]   \\
  3 & SOL2012-May-17T01:25 & M5.1 & 11476 & [926,102]     \\
\hline
\end{tabular}
\end{table}

The microwave emissions during these flares were measured by the
\emph{Nobeyama Radio Polarimeters} \citep[NoRP:][]{Nakajima85} and
\emph{Radioheliograph} \citep[NoRH:][]{Hanaoka94}. The NoRP provides
full-Sun integrated radio fluxes with a temporal cadence of one
second at frequencies of 1\,GHz, 2\,GHz, 3.75\,GHz, 9.4\,GHz,
17\,GHz, and 35\,GHz, while the NoRH gives full-Sun images with a
nominal temporal cadence of one second at frequencies of 17\,GHz and
34\,GHz. Generally, ground-based observations are not as stable as
observations from space, and they often lose some observation data,
in particular for the NoRH images. To acquire temporally uniform
data and compare them with Ly\,$\alpha$ observations, the radio
observations were temporally interpolated to ten-second intervals,
as done for the Ly\,$\alpha$ observations.

\begin{figure}[h!]
\centerline{\includegraphics[width=0.9\textwidth,clip=]{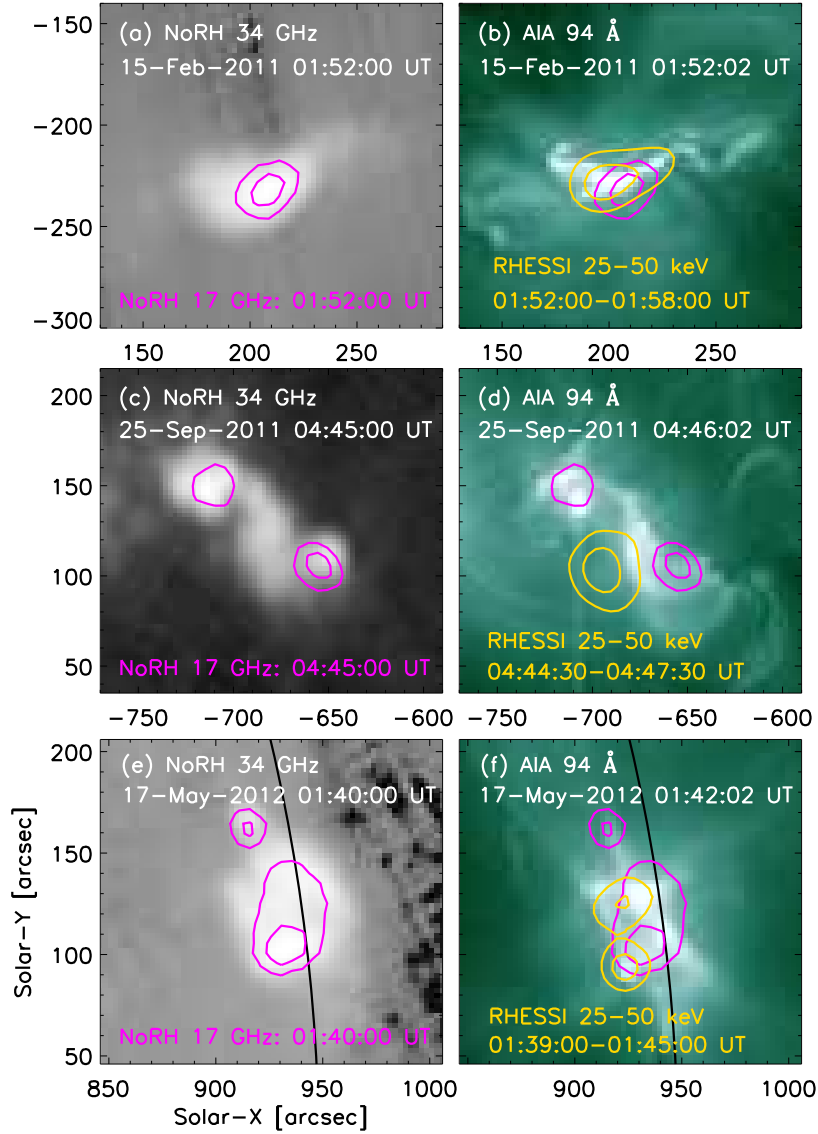}}
\caption{Snapshots at wavelengths of NoRH~34\,GHz (left) and
\emph{Atmospheric Imaging Assembly} (AIA)~94\,{\AA} (right) during
solar flares. The gold and magenta contours represent the
RHESSI~25\,--\,50\,keV and NoRH~17\,GHz emissions, their contour
levels are set at 80\,\% and 50\,\%, respectively. The black curve
in panels e and f marks the solar limb. } \label{images}
\end{figure}

Hard X-ray radiation during these flares were observed by the
\emph{Reuven Ramaty High Energy Solar Spectroscopic Imager}
\cite[RHESSI:][]{Lin02} as well as the \emph{Fermi Gamma-ray Burst
Monitor} \citep[GBM:][]{Meegan09}. RHESSI is a photon-counting
instrument that records events with a precision $\ll$ one second.
However, to obtain images of the Sun in X-rays and $\gamma$-rays,
the instrument was designed to rotate with a period of about four
seconds. This is why the time series of the RHESSI measurements
analyzed here were constructed with a temporal cadence of four
seconds. Note that there are usually some data gaps due to the
RHESSI night times (i.e. the Sun is not visible). \emph{Fermi}/GBM
observes the whole unocculted sky with twelve NaI scintillators
(8\,keV to 1\,MeV) and two bismuth germanate (0.2 to 40\,MeV)
detectors, and thus it offers capabilities for the analyses of not
only $\gamma$-ray bursts but also solar flares. The normal temporal
resolution of \emph{Fermi}/GBM measurements is 0.256\,second, and it
increases to 0.064\,second automatically once a flare starts
\citep{Meegan09}. In the present study, for a better comparison with
observations in Ly\,$\alpha$, the \emph{Fermi}/GBM data were also
reformatted to a temporal cadence of 10.24\,second.

The fast Fourier transformation (FFT) is one of the most powerful
analysis tools, which is widely used in astrophysical and solar
observations to detect QPP signatures
\citep[e.g.][]{Vaughan05,Anfinogentov21}. The quasi-periods in the
present work were determined by analysing the Fourier Power Spectral
Density (PSD) of the time series data. Before applying the FFT, the
light curve is normalized with $(F-F_0)/F_0$, where $F$ is the flux
observed by the above-mentioned instruments, and $F_0$ is the
calculated average flux during the observation time. FFT power
spectrum is usually a superposition of red noise and white noise. So
when determining the significance of possible periodicities in the
PSD, the red noise, an intrinsic property of the observed source due
to erratic, aperiodic brightness changes, has to be accounted for in
order not to severely overestimate the significance of identified
periods \citep{Lachowicz2009}. In observations, the red noise has an
increasing PSD toward longer periods and usually follows a power-law
behavior \citep{Hayes19,Pugh19,Wang20}. For the analysis, the
function $P(f)=Af^{-\alpha}+C$, where $\alpha$ and $A$ represent the
power-law index and normalization parameters, respectively, and $C$
is a constant accounting for the transition between the red and
white noises, is applied to fit the FFT data, and the red noise can
be estimated from the posterior density of the parameters, which is
determined by Markov chain Monte Carlo (MCMC) samples with
Multi-parameter Bayesian inferences. A $\chi^2$ test is used to
assess the significance level of the PSD \citep[see details
in][]{Yuan19,Liang20}. The dominant periods of the detected QPP
periods are considered as the peaks above the significance level,
and their uncertainties are estimated from the full width at half
maximum (FWHM) around the PSD peaks.

\section{Analysis and Results}
\subsection{The X2.2 Flare on 15 February 2011}
On 15 February 2011, a powerful solar flare occurred in the active
region (AR) NOAA~11158, close to the solar center, as shown in
Figure~\ref{images}a and b. The left panels in Figure \ref{images}
show the microwave images at 34\,GHz, with images of 17\,GHz
overplotted as contours. The right panels show the corresponding EUV
images in 94\,{\AA} taken by the \emph{Atmospheric Imaging Assembly}
\citep[AIA:][]{Lemen2012}  onboard the \emph{Solar Dynamics
Observatory} (SDO), overplotted with image contours taken at 17\,GHz
(magenta) and 25\,--\,50\,keV (yellow).

\begin{figure}
\centering
\includegraphics[width=\linewidth,clip=]{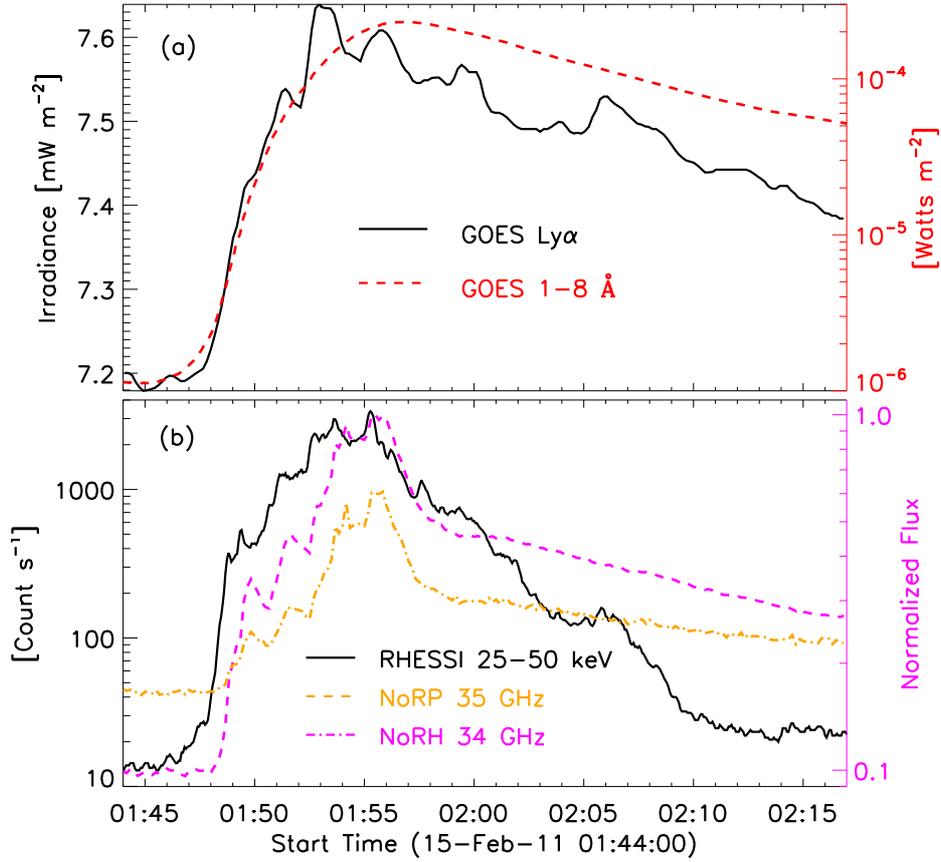}
\caption{Light curves of the solar flare on 15 February 2011. Panel
a: Full-disk light curves at wavelengths of Ly\,$\alpha$ (black) and
GOES~1\,--\,8\,{\AA} (red). Panel b: Full-disk light curves in
RHESSI~25\,--\,50\,keV (black), NoRP~35\,GHz (orange), as well as
the locally integrated  flux at NoRH 34\,GHz (magenta) over the
flare region shown in Figure~\ref{images}a. Note that the
discontinuities in the RHESSI light curves, caused by the RHESSI
attenuator changes, have been empirically corrected.}
\label{f110215}
\end{figure}

Figure \ref{f110215} shows the full-Sun integrated light curves in
different wavelengths. In panel a, the red line shows the SXR
emission in 1\,--\,8 {\AA}, according to which the flare started at
01:44 UT, peaked at 01:56 UT, and then underwent a long decay phase.
It is the first X-class (X2.2) flare in Solar Cycle 24 and has been
extensively studied. The black line in Figure \ref{f110215}a shows
the corresponding Ly\,$\alpha$ emission, which exhibits several
regular and periodic pulses (signatures of QPPs during the solar
flare). The corresponding HXR and microwave emissions that are
supposed to have a nonthermal origin \citep{Holt69,Saint05} are
shown in Figure \ref{f110215}b, where the black line shows the HXR
flux in 25\,--\,50\,keV from RHESSI and the orange line shows the
microwave emission at 35\,GHz from NoRP. As a comparison, the
locally integrated microwave flux over the flare region, calculated
from NoRH images at 34\,GHz, was also overplotted (magenta line).
Note that the discontinuities in the RHESSI light curve, caused by
the RHESSI attenuator changes, have been empirically corrected, and
the microwave light curves were normalized to their maximum values
and shifted vertically to avoid overlap. Similar to the Ly\,$\alpha$
emission, all these nonthermal fluxes are characterized by a series
of small pules.

\begin{figure}
\centering
\includegraphics[width=\linewidth,clip=]{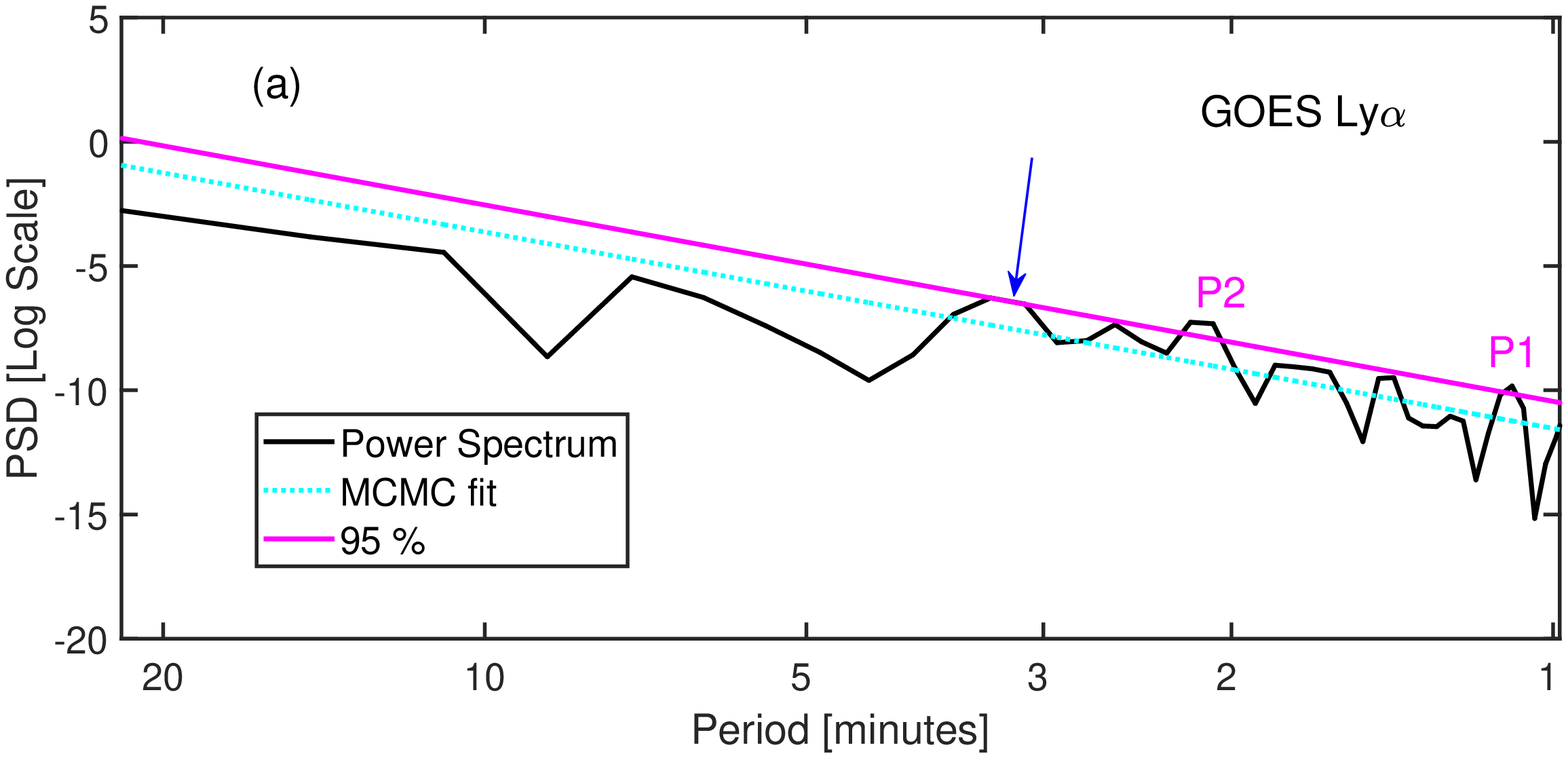}
\includegraphics[width=\linewidth,clip=]{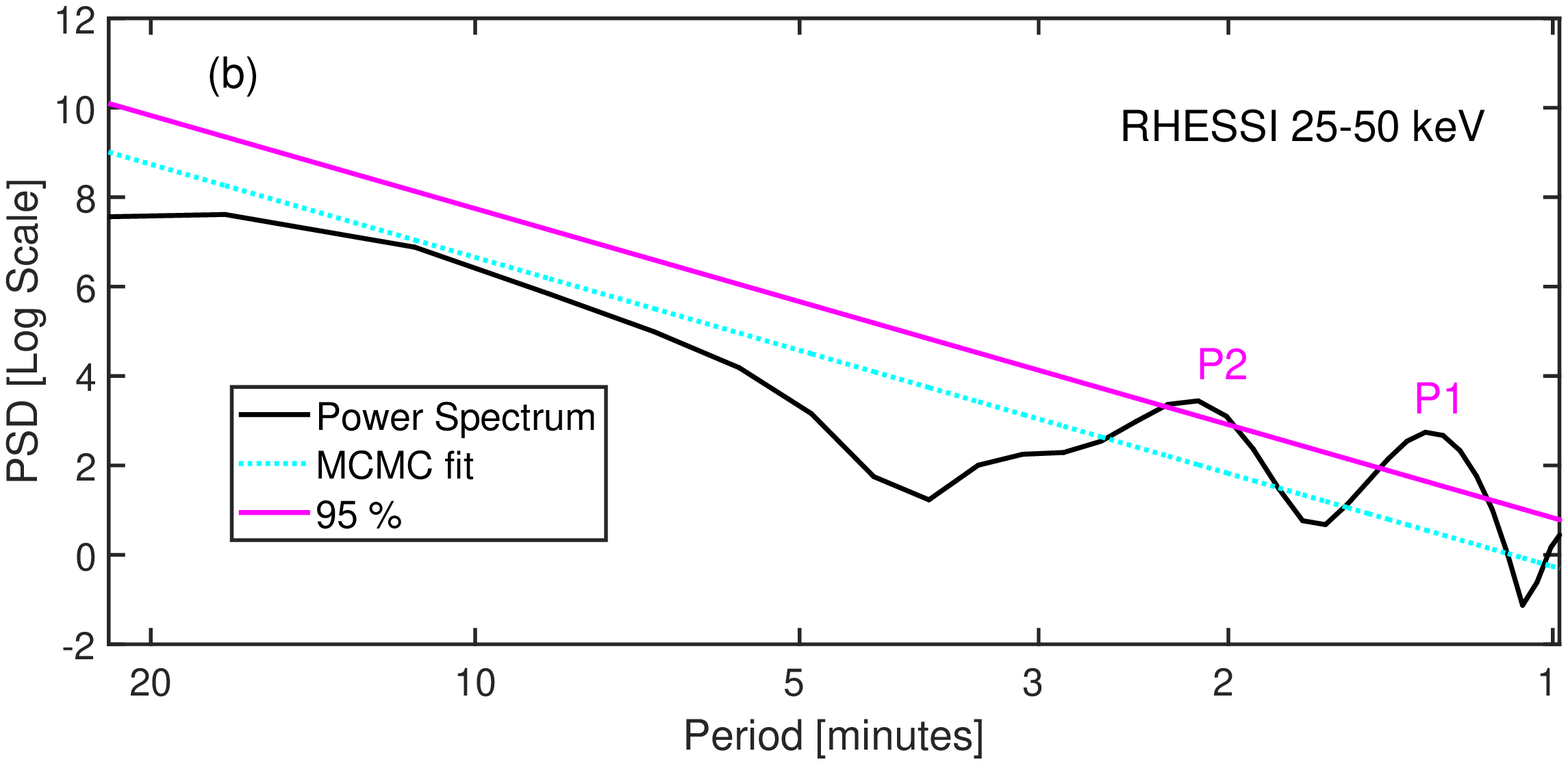}
\caption{PSDs for Ly\,$\alpha$ and HXR emissions of the X2.2 flare
in log--log space. The cyan line is the best (MCMC) fit, and the
magenta line represents the 95\,\% confidence level. The double
periods studied in this work are marked with ``P1'' and ``P2'',
while the blue arrow indicates a quasi-period of roughly three
minutes.} \label{psd1}
\end{figure}

Figure~\ref{psd1} presents the QPP analysis of the X2.2 flare at
wave channels of Ly\,$\alpha$ (panel a) and HXR~25\,--\,50\,keV
(panel b) in log--log space. The time interval used to calculate the
PSD is the full flare duration, as shown in Figure~\ref{f110215}.
The cyan line represents the best MCMC fit and the magenta line
indicates the 95\,\% confidence (or 5\,\% significance) level, which
is often used to detect the solar QPPs
\citep[e.g.][]{Pugh17,Kolotkov18,Yuan19,Li20b,Liang20}. In Figure
\ref{psd1}a, there are two significant peaks beyond the confidence
level, which are considered as double-period QPPs. The double
periods are estimated to be $\approx$1.09~minutes (``P1'') and
$\approx$2.15~minutes (``P2''), the ratio of which is equal to 1.97,
which are listed in Table \ref{tab2}. There is also a weak peak that
is just reaching the confidence level, as indicated by the blue
arrow. It reveals a period value of three minutes, which agrees well
with the three-minute oscillations reported by \cite{Milligan17} for
the same flare.

\begin{figure}
\centering
\includegraphics[width=\linewidth,clip=]{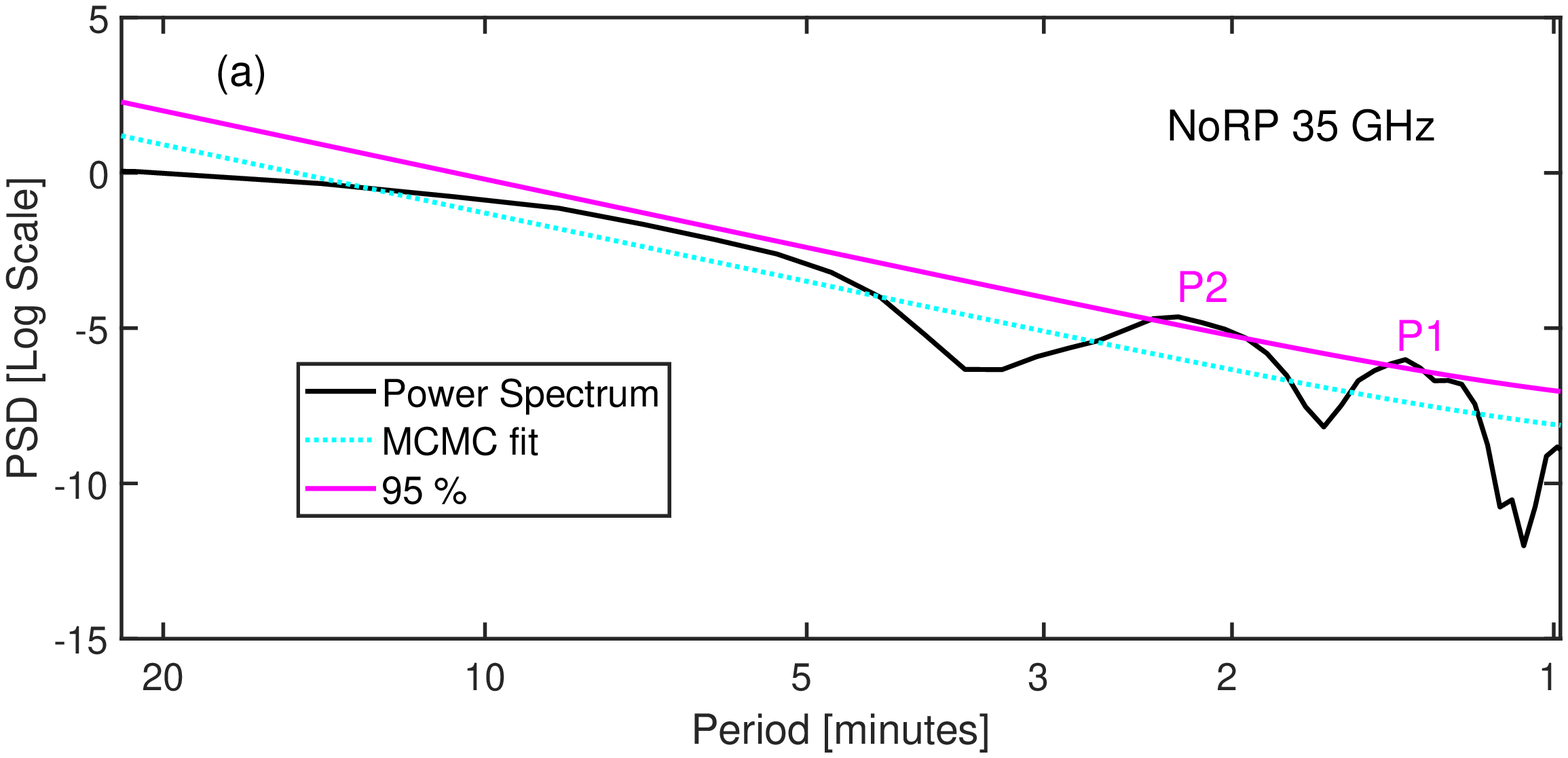}
\includegraphics[width=\linewidth,clip=]{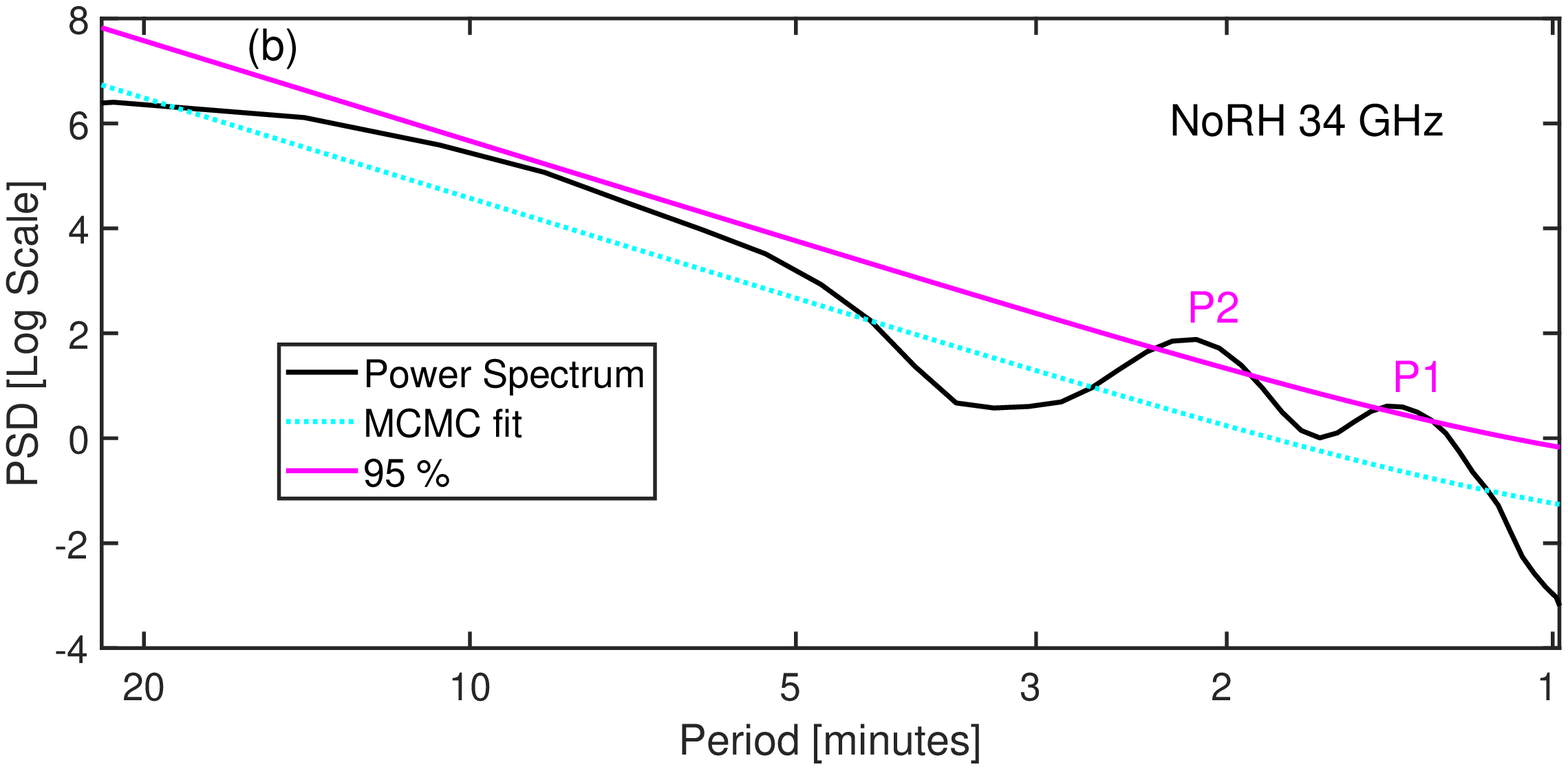}
\caption{PSDs for microwave emissions of the X2.2 flare in
log--log space. The cyan line is the best (MCMC) fit, and the magenta
line represents the 95\,\% confidence level. The double periods
studied here are marked with ``P1'' and ``P2''.} \label{psd2}
\end{figure}

In order to verify the oscillation periods, we did the same QPP
analysis for the contemporaneous observations in other wavelengths.
Figure \ref{psd1}b and Figure~\ref{psd2} show the PSDs calculated
from the light curves in HXR 25\,--\,50\,keV and microwaves,
respectively. Note that, the PSD in Figure~\ref{psd2}a is calculated
from the full-Sun microwave emission at 35\,GHz from NoRP, while the
PSD in Figure~\ref{psd2}b is calculated from the local microwave
emission (integrated over the flare region) at 34\,GHz from NoRH.
They both exhibit nearly the same double periods (``P1'' and ``P2'')
as those in HXR 25\,--\,50\,keV (as listed in Table \ref{tab2}),
implying their nonthermal origin from the flare region. Comparing
with double periods in the Ly\,$\alpha$ emission, ``P2'' is quite
similar, while ``P1'' is a little different, for instance, a
deviation of about 0.28~minutes is detected between the Ly\,$\alpha$
and microwave emissions, as seen in Table \ref{tab2}. On the other
hand, the PSD peaks are very broad in the nonthermal emission at HXR
and microwave channels, suggesting quite large uncertainties of
HXR/microwave QPPs.

\subsection{The M7.4 Flare on 25 September 2011}
On 25 September 2011, the GOES/XRS recorded a soft X-ray event that
started at 04:31 UT, had its maximum at 04:50 UT, and ended at 05:05
UT. The event was classified as an M7.4 flare according to the peak
soft X-ray emission. The imaging observations reveal that it took
place in AR 11302, as shown in Figure~\ref{images}c and d.

\begin{figure}
\centering
\includegraphics[width=\linewidth,clip=]{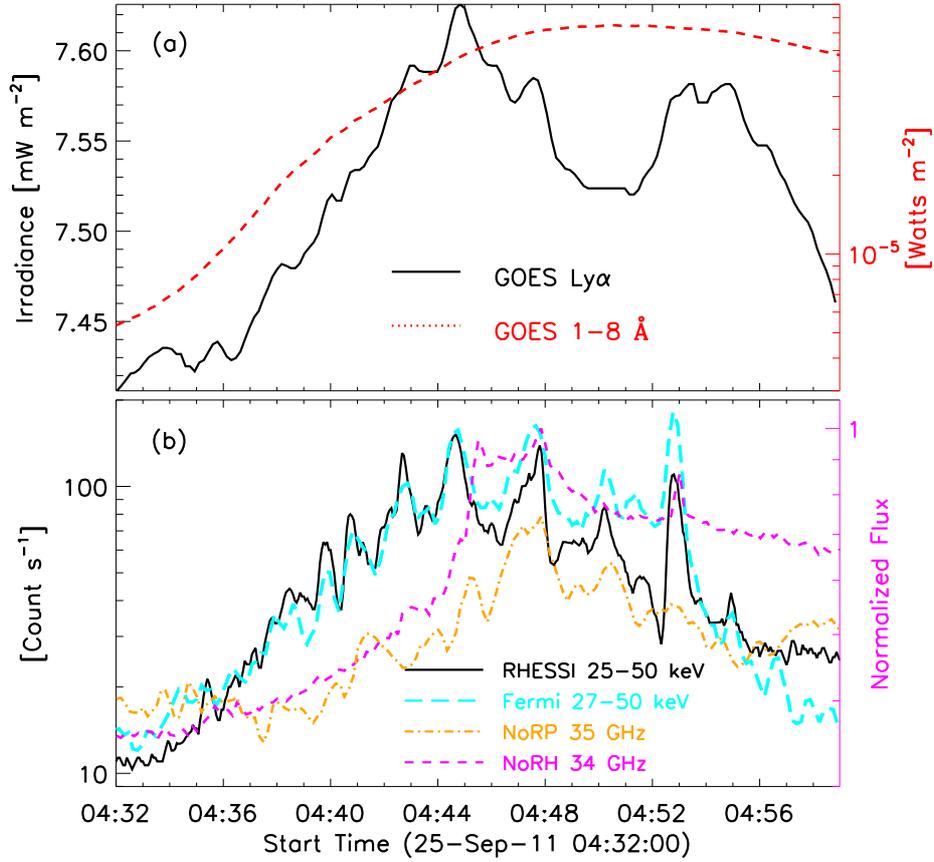}
\caption{Light curves of the solar flare on  25 September 2011.
Panel~a: Full-disk light curves at wavelengths of Ly\,$\alpha$
(black), and GOES~1\,--\,8\,{\AA} (red). Panel~b: Full-disk light
curves in RHESSI~25\,--\,50\,keV (black) and
\emph{Fermi}~27\,--\,50\,keV (cyan) and NoRP~35\,GHz (orange), as
well as the locally integrated microwave flux at NoRH~34\,GHz
(magenta) over the flare region shown in Figure~\ref{images}b. Here
the \emph{Fermi} data are from the GBM detetor NaI 4 and the
discontinuities in the RHESSI count rate, caused by the RHESSI
attenuator changes, have been empirically corrected.}
\label{f110925}
\end{figure}

\begin{figure}
\centering
\includegraphics[width=\linewidth,clip=]{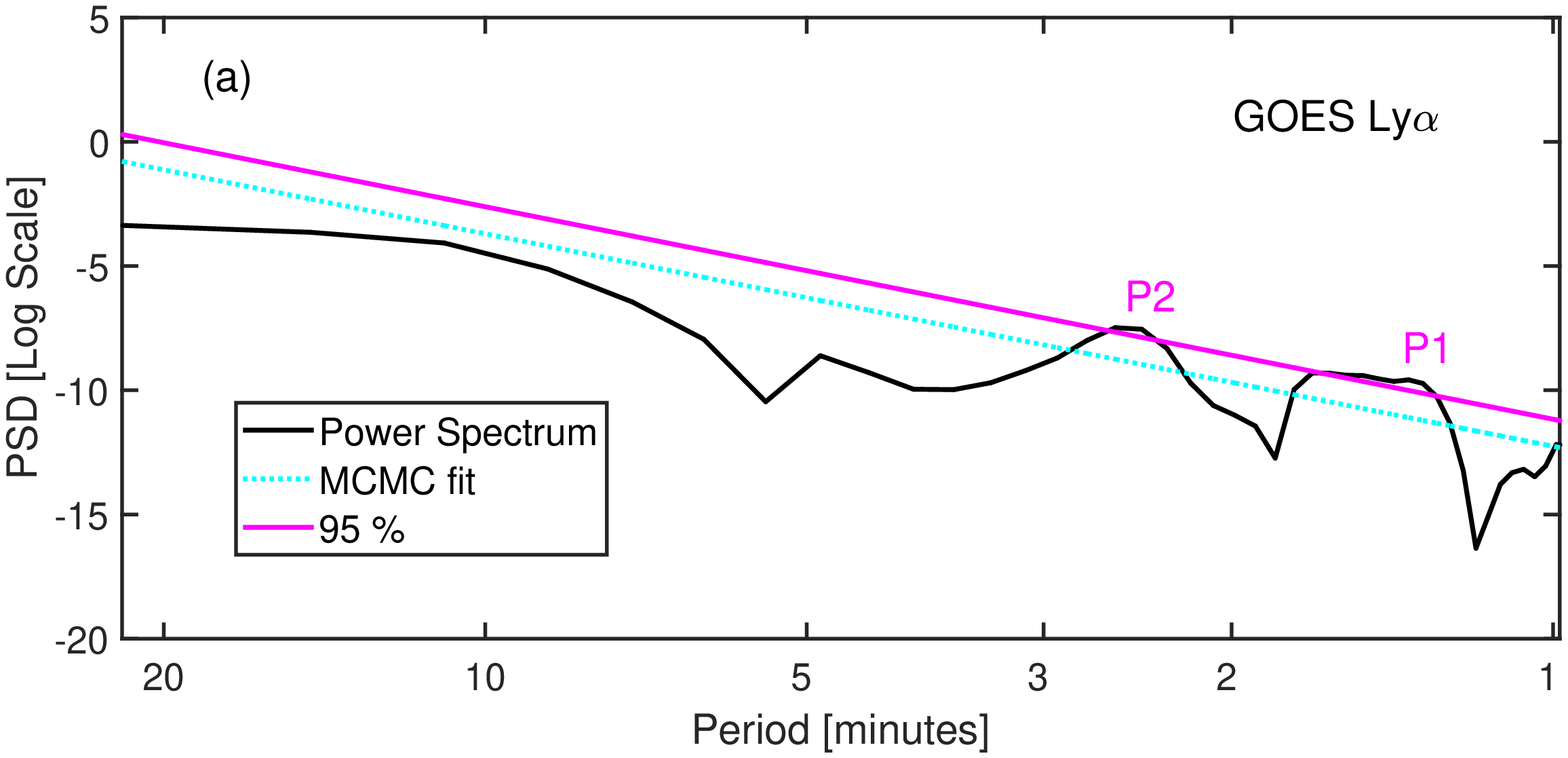}
\includegraphics[width=\linewidth,clip=]{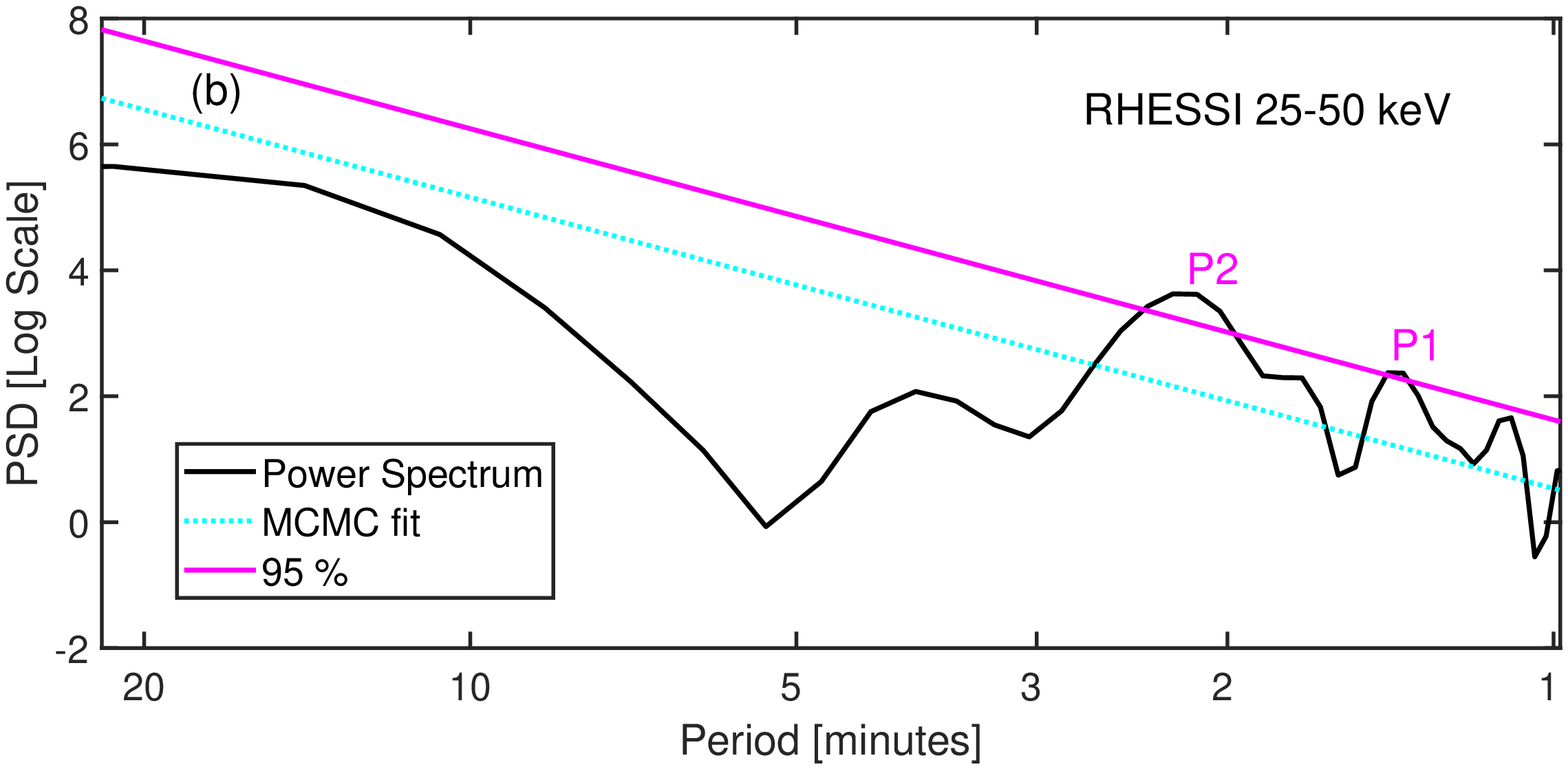}
\includegraphics[width=\linewidth,clip=]{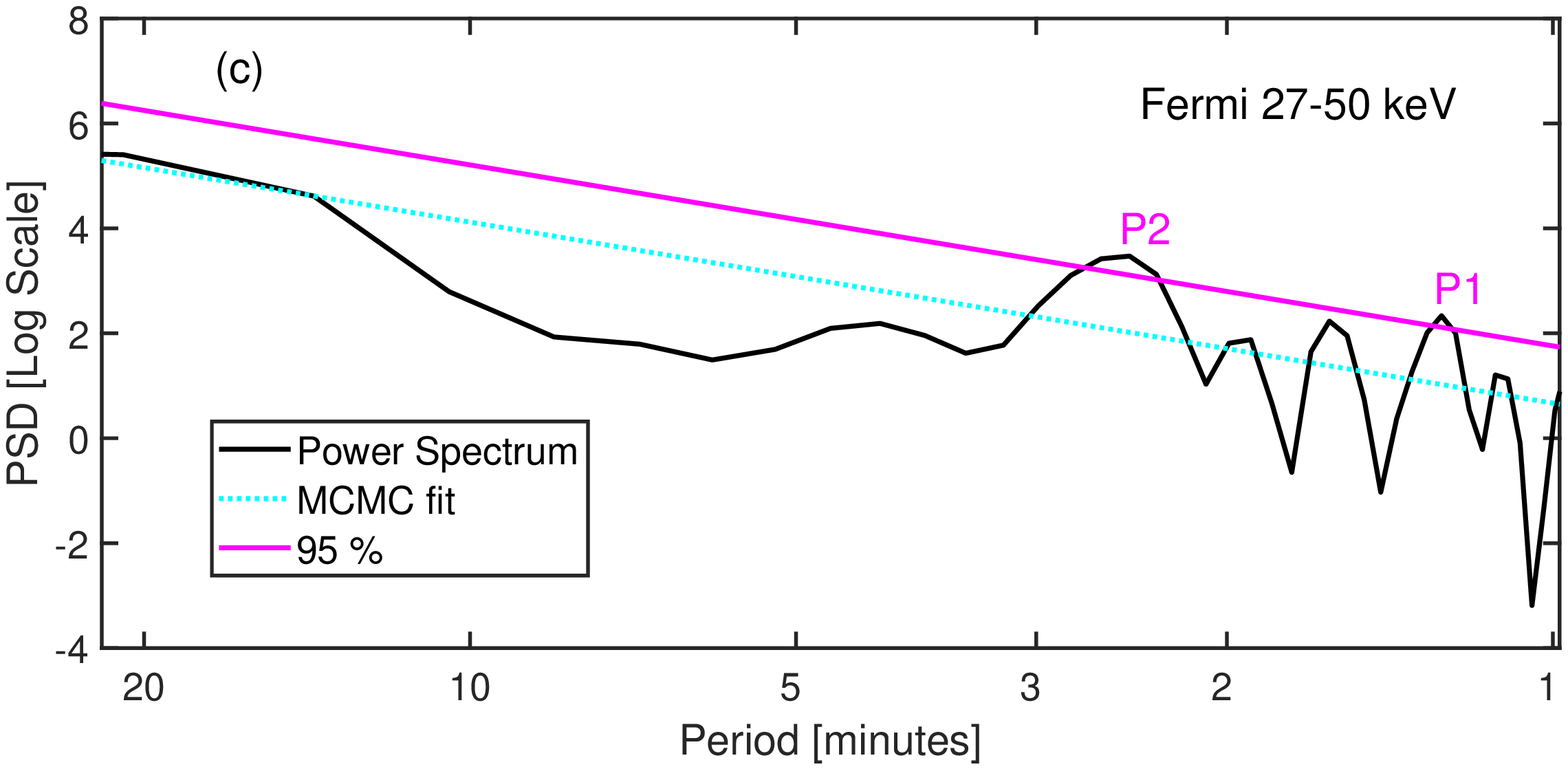}
\caption{PSDs for Ly\,$\alpha$ and HXR emissions of the M7.4 flare
in log--log space. The cyan line is the best (MCMC) fit, and the
magenta line represents the 95\,\% confidence level. The double
periods studied here are marked with ``P1'' and ``P2''.}
\label{psd3}
\end{figure}

\begin{figure}
\centering
\includegraphics[width=\linewidth,clip=]{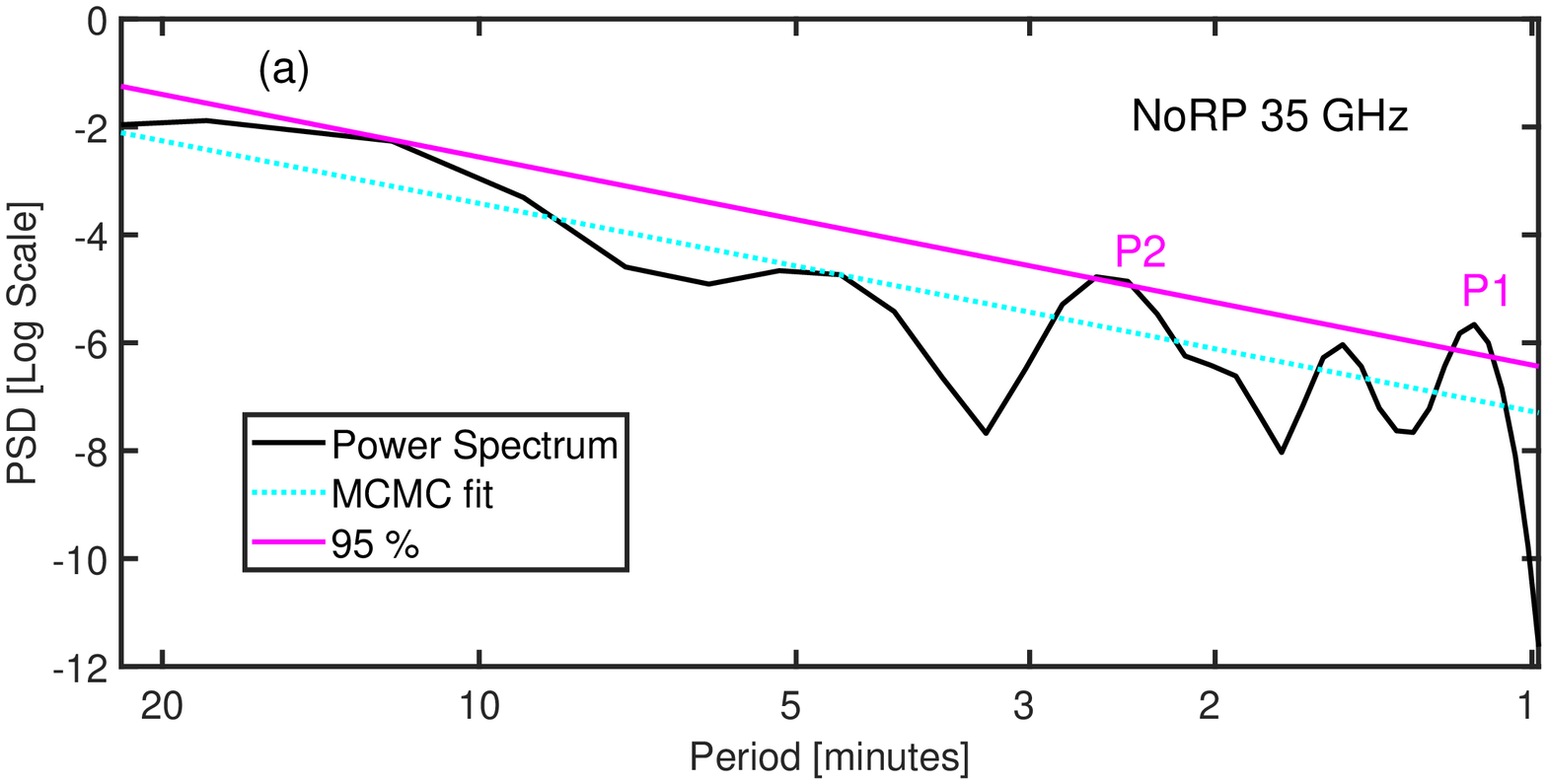}
\includegraphics[width=\linewidth,clip=]{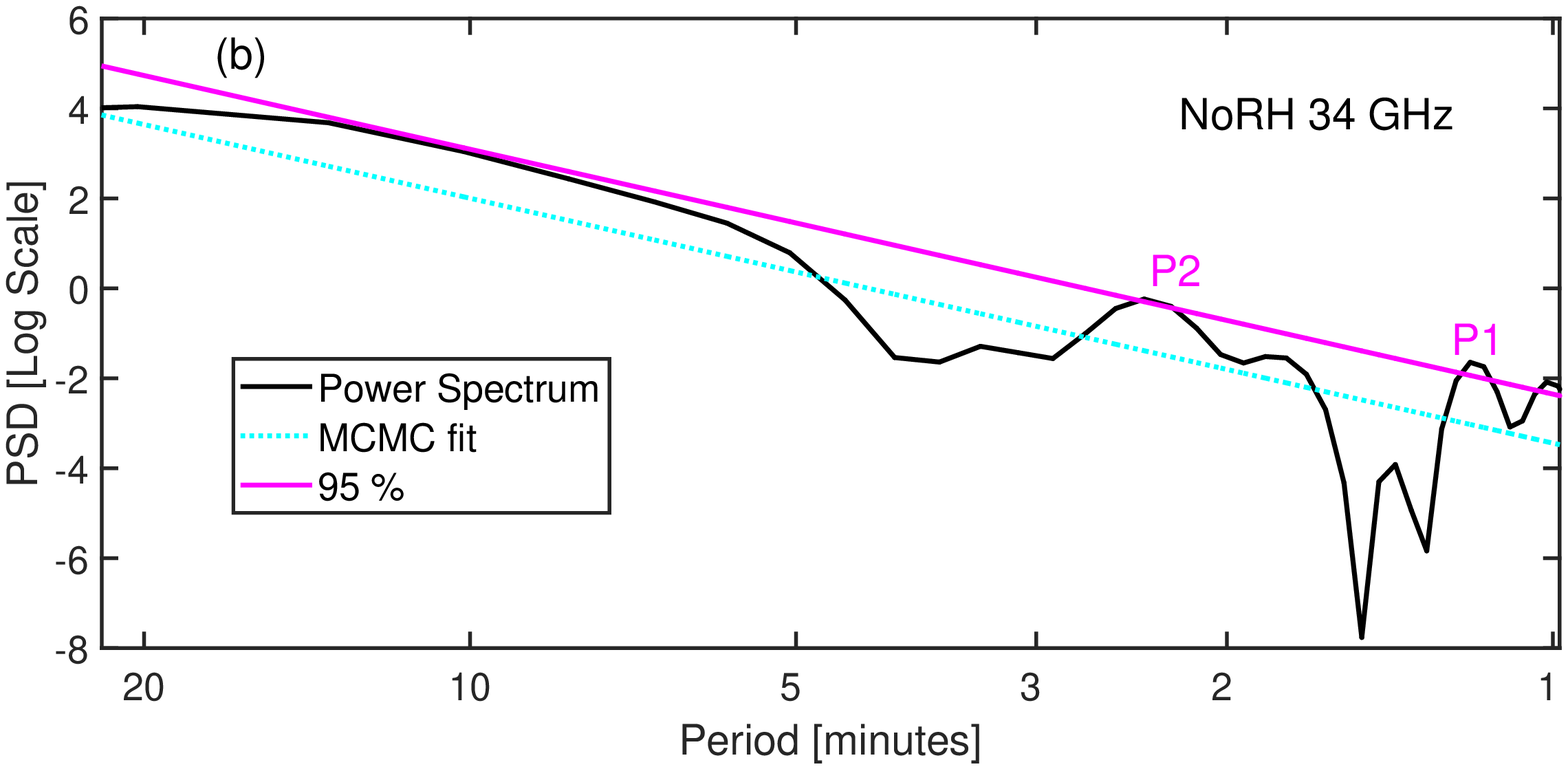}
\caption{PSDs for microwave emissions of the M7.4 flare in log--log
space. The cyan line is the best (MCMC) fit, while the magenta line
represents the 95\,\% confidence level. The double periods studied
here are marked with ``P1'' and ``P2''.} \label{psd4}
\end{figure}

Figure~\ref{f110925} shows light curves of the flare in different
wavelengths between 04:32~UT and 04:59~UT. Panel a gives light
curves of full-Sun integrated solar radiation in Ly\,$\alpha$ line
(black) and SXR~1\,--\,8\,{\AA} (red). As can be seen, there are
multiple regular and periodic pulses in the Ly\,$\alpha$ light
curve, which are supposed to be signatures of QPPs during the flare.
Panel b gives the full-Sun integrated solar radiations at channels
of RHESSI~25\,--\,50\,keV (black), \emph{Fermi}~27\,--\,50\,keV
(cyan), and NoRP~35\,GHz (orange), as well as the locally integrated
microwave flux over the flare region at NoRH~34\,GHz (magenta). For
the analysis, the CTIME data from the \emph{Fermi}/GBM detector NaI
4 (Sun-facing) is used, and the discontinuities in the RHESSI light
curve, caused by the RHESSI attenuator changes, have been
empirically corrected. All of these solar emissions are supposed to
have a nonthermal origin, which exhibit similar QPP signatures as
those in Ly\,$\alpha$ light curve. The similarity between the RHESSI
and \emph{Fermi} measurements rules out the possibility of an
instrumental effect.

To find the periodicity of QPPs, we calculate the PSDs of the light
curves shown in Figure \ref{f110925}. Our calculation results are
shown in Figures~\ref{psd3} and \ref{psd4}. From the PSD of the
Ly\,$\alpha$ light curve (Figure~\ref{psd3}a), two broad peaks
(labeled ``P1'' and ``P2'') with  periodicities of
$\approx$1.33\,minutes and $\approx$2.42\,minutes stand out above
the 95\,\% confidence level. The ratio between the double periods
was estimated to be 1.82. The double-period oscillations were also
detected in the full-Sun HXR fluxes measured by the
RHESSI~25\,--\,50\,keV (Figure~\ref{psd3}b) and
\emph{Fermi}~27\,--\,50\,keV (Figure~\ref{psd3}c), as well as the
full-Sun and locally integrated microwave fluxes at the frequencies
of NoRP~35\,GHz and NoRH~34\,GHz (see Figure~\ref{psd4}). The
quasi-periods are listed in Table \ref{tab2}. As can be seen, the
flare emission in the Ly\,$\alpha$, HXR and microwave show a very
similar long period (``P2'') except for the RHESSI 25\,--\,50\,keV
flux, which exhibits a slightly shorter period. For the short period
``P1'', it is also slightly different; for instance, the flare
emission in the Ly\,$\alpha$ and HXR show a quite similar period
(``P1''), while the period in the microwave emission is slightly
shorter.

\subsection{The M5.1 Flare on 17 May 2012}

\begin{figure}
\centering
\includegraphics[width=\linewidth,clip=]{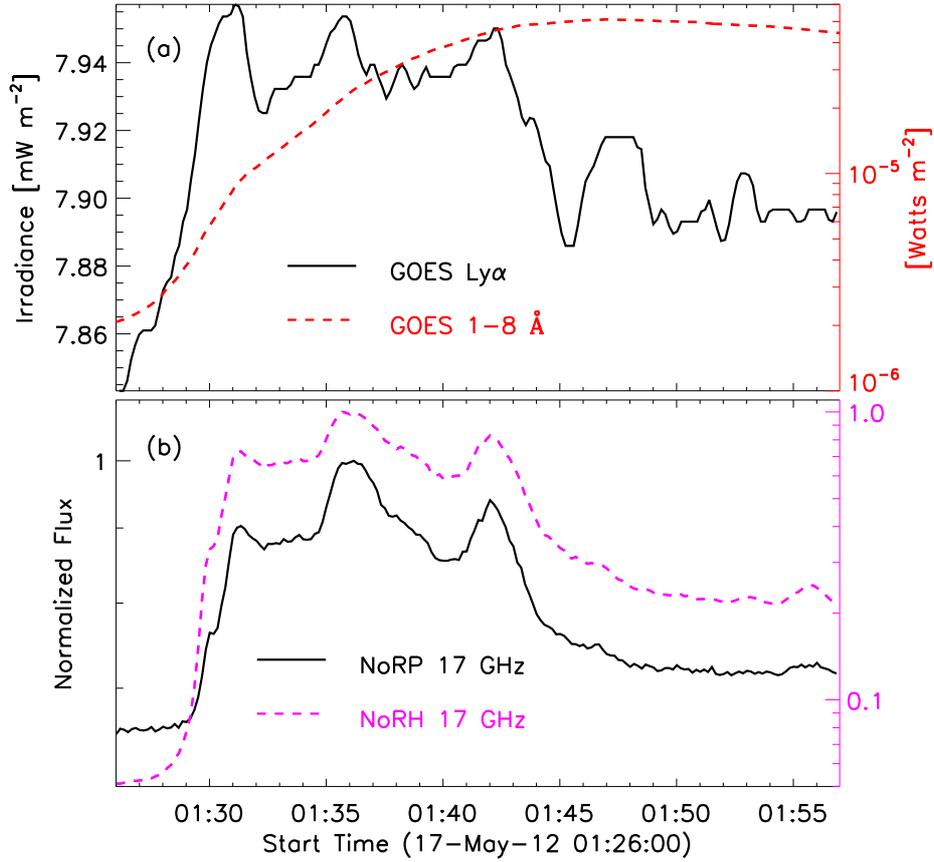}
\caption{Light curves of the solar flare on  17 May 2012. Panel~a:
Full-disk light curves at wavelengths of Ly\,$\alpha$ (black) and
GOES~1\,--\,8\,{\AA} (red). Panel~b: Full-disk light curves in
NoRP~17\,GHz (black), and the local NoRH flux at the frequency of
microwave~17\,GHz (magenta) that is integrated over the flare region,
as shown in Figure~\ref{images}c and d.}  \label{f120517}
\end{figure}

\begin{figure}
\centering
\includegraphics[width=\linewidth,clip=]{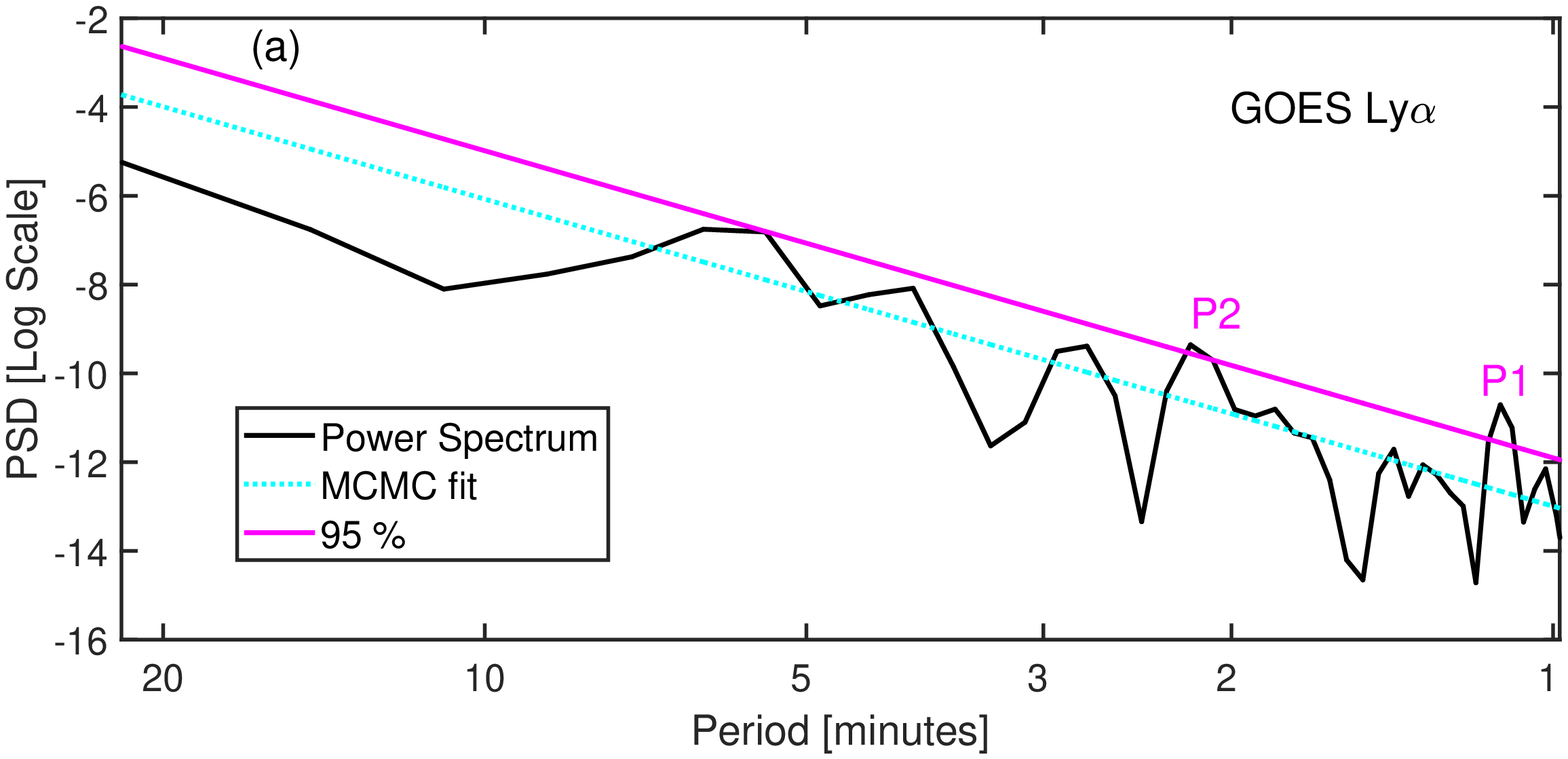}
\includegraphics[width=\linewidth,clip=]{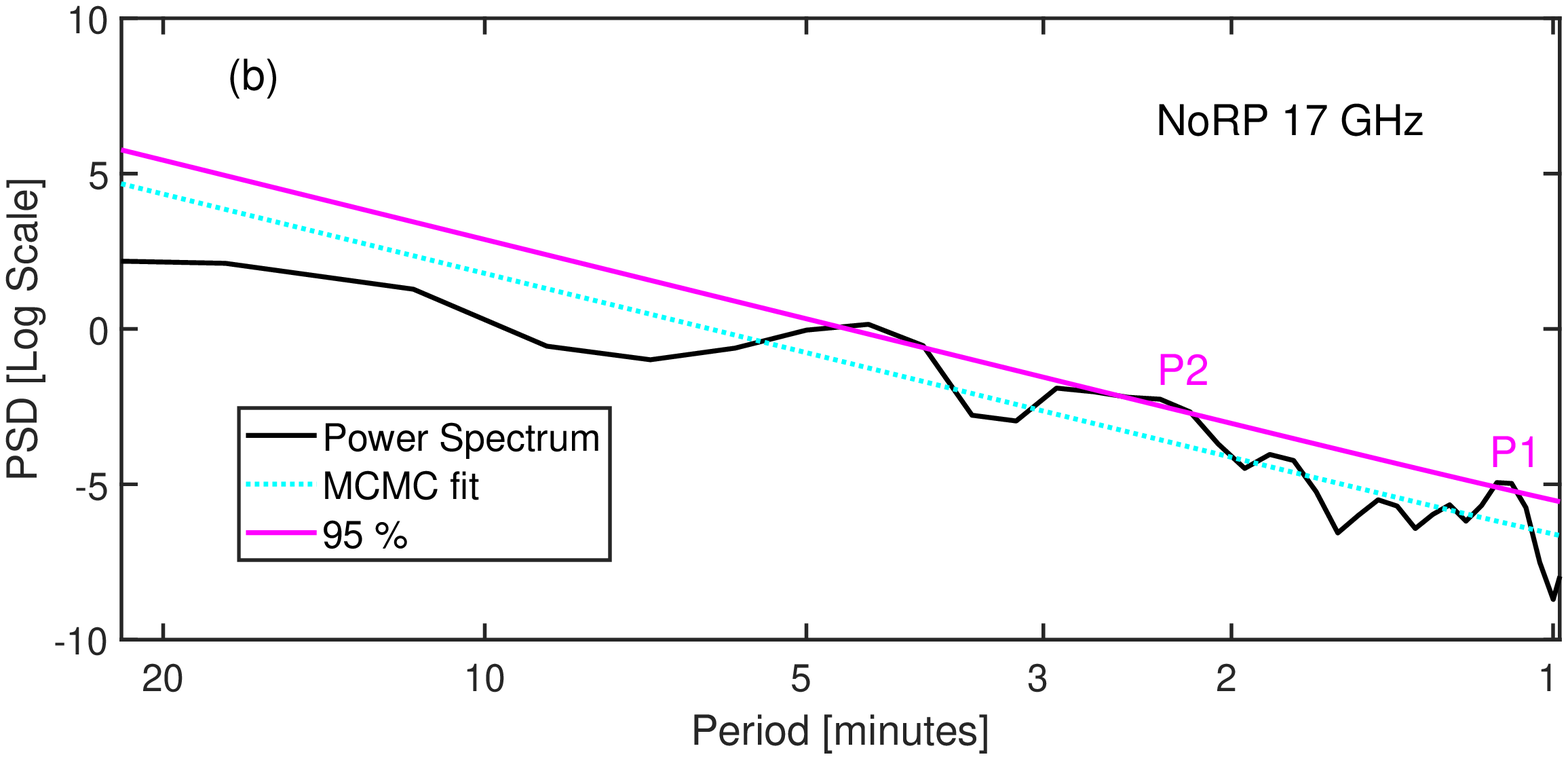}
\includegraphics[width=\linewidth,clip=]{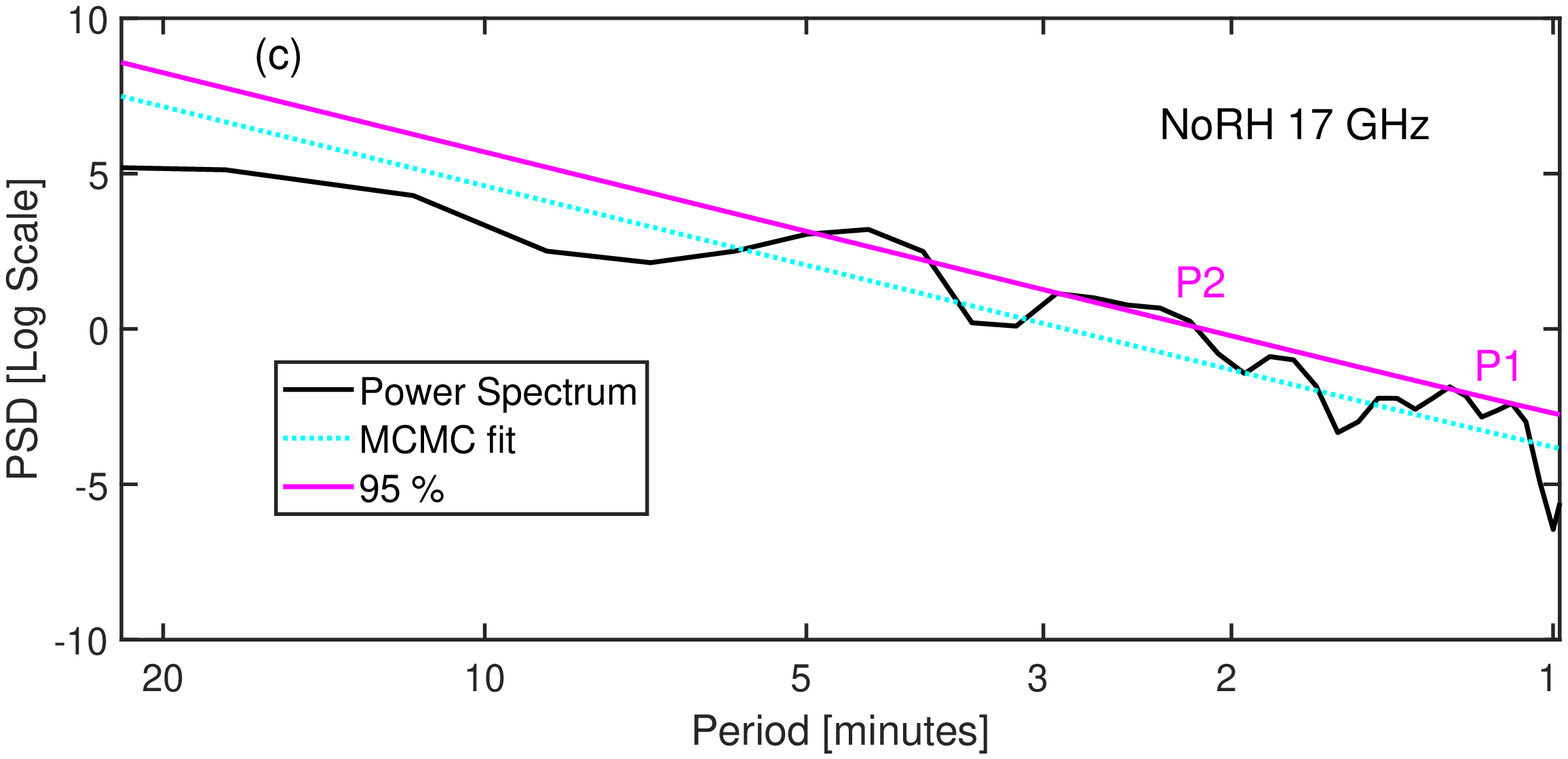}
\caption{PSDs for Ly\,$\alpha$ and microwave emissions of the M5.1
flare in log--log space. The cyan line is the best (MCMC) fit,
and the magenta line represents the 95\,\% confidence level. The
double periods studied here are marked with ``P1'' and
``P2''.} \label{psd5}
\end{figure}

On 25 May 2012, a (west) limb flare occurred in  AR 11476, as shown
in Figure~\ref{images}~e and f. According to the GOES observations
in 1\,--\,8\,{\AA}, the flare was  classified as an M5.1 flare,
which started at 01:25 UT, peaked at 01:47 UT, and ended at 02:14
UT. Figure~\ref{f120517} draws the light curves for this flare
between about 01:26~UT and 01:57~UT. Similar to the above-described
two flares, Figure~\ref{f120517}a presents light curves of the
full-Sun radiations in Ly\,$\alpha$ (black) and SXR~1\,--\,8\,{\AA}
(red), and Figure~\ref{f120517}b displays the full-Sun (black) and
locally integrated (magenta) microwave emissions at the frequency of
17\,GHz, recorded by NoRP and NoRH, respectively. As can be seen,
the light curves in both Ly\,$\alpha$ and microwave emissions show
clear signatures of QPP features. Note that RHESSI was in the orbit
night until 01:39 UT and images of the NoRH 34\,GHz exhibit a very
poor quality; thus observations from these two instruments were not
used here.

Figure~\ref{psd5} presents the PSDs of light curves at wavelengths
of Ly\,$\alpha$ (panel a) and 17\,GHz (panels b and c) in log--log
space. Note that, Figure~\ref{psd5}b shows the PSDs of the full-Sun
integrated light curve at 17\,GHz while Figure~\ref{psd5}c shows the
PSDs of the locally integrated light curve. Above the 95\,\%
confidence level, we found three significant peaks with dominant
periods of $\approx$5\,minutes, $\approx$2.19\,minutes, and
$\approx$1.12\,minutes in the Ly\,$\alpha$ emission. The latter two
periods and the ratio (i.e., 1.96) between them are quite similar to
the double periods found in previous two flares, which may imply a
similar physical process. On the other hand, similar double periods
can also be found in the microwave emission. Regarding the
five-minute period, it was supposed to result from the three main
pulses during the impulsive phase between $\approx$01:29~UT and
$\approx$01:44~UT (see Figure~\ref{f120517}), which may
corresponding to three repeated magnetic reconnections, but it is
beyond the scope of this study.

\section{Conclusions and Discussion}
The Ly\,$\alpha$ line is the brightest emission line in the solar UV
spectrum, but it has  rarely been studied due to limited
observations, in particular the imaging observations. In the present
study, we investigated QPPs observed in Ly\,$\alpha$, HXR, and
microwave emissions during three powerful solar flares that occurred
on 15 February 2011, 25 September 2011, and 17 May 2012,
respectively. By applying the MCMC sampling technique, QPPs with
double periods of about two\,minutes (P2) and one\,minute (P1) were
first detected in the Ly\,$\alpha$ emission recorded by the
GOES/EUVS. Note that the short period (P1) actually ranges between
about 1.09 and 1.33\,minutes, and the long period (P2) varies from
roughly 2.15 to 2.42\,minutes; here we simply regard them as one-
and two-minute QPPs, as was done in previous studies
\citep{Milligan17,Ning17,Li20c}. Then using the same technique, the
flare QPPs with double periods were also found from light curves of
full-Sun HXR (25\,--\,50\,keV, 27\,--\,50\,keV) and microwave
(35/17\,GHz) emissions, with the HXR emission recorded by RHESSI and
\emph{Fermi}/GBM, and microwave emission recorded by  NoRP. As a
comparison, the QPP analysis technique was also applied to local
light curves integrated over the flaring region in NoRH images
(34/17\,GHz), and again the double periods were obtained. This
implies that the QPPs originate from the flaring region rather than
the solar background radiation. Table \ref{tab2} shows the double
periods as well as the ratio between them for the three flares under
study.

\begin{table}[]
\centering \setlength{\tabcolsep}{4pt}
\caption{Double periods observed in the Ly\,$\alpha$, HXR, and microwave emissions during solar flares.}
\begin{tabular}{l l c c c}
\hline
\multicolumn{2}{l}{} & P1 [minutes] & P2 [minutes] & Ratio (P2/P1)  \\
\multirow{4}{*}{SOL2011-02-15T01:44}
& GOES Ly\,$\alpha$  & 1.09$\pm$0.05 & 2.15$\pm$0.15 & 1.97$\pm$0.23   \\
& RHESSI 25\,--\,50\,keV  & 1.26$\pm$0.19 & 2.13$\pm$0.32 & 1.69$\pm$0.52  \\
& NoRP 35\,GHz & 1.37$\pm$0.12 & 2.17$\pm$0.36 & 1.58$\pm$0.41  \\
& NoRH 34\,GHz  & 1.37$\pm$0.19 & 2.13$\pm$0.33 & 1.56$\pm$0.47  \\
\multirow{5}{*}{SOL2011-09-25T04:31}
& GOES Ly\,$\alpha$ & 1.33$\pm$0.19  & 2.42$\pm$0.24 & 1.82$\pm$0.45  \\
& RHESSI 25\,--\,50\,keV  & 1.37$\pm$0.16 & 2.24$\pm$0.31 & 1.64$\pm$0.43  \\
& \emph{Fermi} 27\,--\,50\,keV  & 1.27$\pm$0.12 & 2.46$\pm$0.33 & 1.94$\pm$0.45  \\
& NoRP 35\,GHz  & 1.12$\pm$0.11 & 2.41$\pm$0.29 & 2.15$\pm$0.48  \\
& NoRH 34\,GHz  & 1.19$\pm$0.10 & 2.39$\pm$0.28 & 2.00$\pm$0.41 \\
\multirow{4}{*}{SOL2012-05-17T01:25}
& GOES Ly\,$\alpha$ & 1.12$\pm$0.05 & 2.19$\pm$0.16 & 1.96$\pm$0.23  \\
& NoRP 17\,GH & 1.12$\pm$0.06 & 2.33$\pm$0.35 & 2.08$\pm$0.43  \\
& NoRH 17\,GH  & 1.25$\pm$0.09 & 2.33$\pm$0.36 & 1.86$\pm$0.43  \\
\hline
\end{tabular}
\label{tab2}
\end{table}

With regard to the X2.2 flare on  15 February 2011, it has been
studied by many authors \citep[see][for a simple
statistic]{Milligan18}. Using the \emph{Chinese Solar Broadband
Radio Spectrometer} at Huairou, \cite{Tan12} found QPPs with a
period of $\approx$375\,ms in the microwave emission at 6.8\,GHz and
attributed it to a very short period pulsation (VSP). Then,
\cite{Milligan17} reported a three-minute QPP in the full-Sun
Ly\,$\alpha$ and Lyman continuum emissions (similar QPP can also be
seen in our analysis result, as indicated by the blue arrow in
Figure~\ref{psd1}a), and explained it as the acoustic wave in the
chromosphere. \cite{Milligan17} also found a period of about two
minutes in both HXR and chromospheric emission, which is consistent
with the two-minute QPP reported here. It should be stated that,
instead of simply applying the wavelet-analysis technique to a
detrended light curve (having removed the filtered time profile) as
was done by \cite{Milligan17}, we performed the FFT analysis for
light curves without detrending applied, thus more oscillation
periods are detected, such as the one-minute QPP,  which has been
extensively studied at wavelengths of X-ray and EUV during different
solar flares \citep[see][]{Van11,Cho16,Ning17,Hayes19,Li20c}. Our
findings here agree with previous results, suggesting that the QPP
analysis technique \citep{Liang20} that we used is reliable.

The double or even more periods of flare QPPs were previously found
in X-ray emission \citep{Zimovets10}, H$\alpha$ images
\citep{Srivastava08}, microwave emission \citep{Tan10}, and  even UV
spectral lines \citep{Tian16}. However, these periods are usually
detected in different phases of solar flares, for example one period
in the impulsive phase and the other period in the decay phase
\citep{Huang14,Kolotkov18,Hayes19}. Moreover, the ratio between
these periods is usually found to be significantly different from
two \citep[e.g.][]{Srivastava08,Inglis09,Tian16,Hayes19}. In the
present work, double periods of QPPs were observed during three
flares in both Ly\,$\alpha$ and nonthermal (HXR/microwave)
emissions. The period ratio detected in the Ly\,$\alpha$ emission
was very close to two, which is consistent with the theoretical
expectation between the fundamental and harmonic modes for the
weakly dispersive MHD modes
\citep[e.g.,][]{Nakariakov09,Nakariakov20}. However, considering
that the PSDs were calculated from the full flare duration, we can
not rule out the possibility that   the double periods occurred at
different phases of the flares. The period ratios in the HXR and
microwave emissions were found to vary from about 1.56 to 2.15, some
of which significantly deviated from the theoretical expectation,
i.e., for the highly dispersive modes. On the other hand,
considering the large uncertainties in the HXR and microwave
emissions, the period ratios here are roughly equal to that found in
the Ly\,$\alpha$ emission for the same flare event.

Flare QPPs observed in HXR and microwaves are supposed to be
associated with periodic acceleration of charged particles that
radiate HXR and microwave emissions via bremsstrahlung and
gyrosychrotron mechanisms, respectively
\citep{Dulk85,Aschwanden87,Masuda94,Druett19}, and they are often
well correlated. For example, the kink oscillation excited in a
coronal loop near the reconnection site could periodically trigger
magnetic reconnections, during which charged particles are
periodically accelerated, inducing QPPs in HXR and microwave
\citep{Foullon05,Nakariakov06,Inglis09}. The Ly\,$\alpha$ emission
enhancements during solar flares were found to be highly synchronous
with emission enhancements in HXR \citep{Nusinov06,Rubio09}, which
suggests that the same population of  energetic electrons is
responsible for radiation in Ly\,$\alpha$. Thus we suppose that QPPs
in Ly\,$\alpha$ are also associated with nonthermal electrons that
are accelerated by repetitive magnetic reconnections during solar
flares. The repeated magnetic reconnections that generate the long
period (P2) might be spontaneous via magnetic dripping
\citep{Li20c}, or modulated by a MHD mode, such as the standing
sausage mode \citep{Van11}. The observation of double-periodic QPPs
on the Sun can be utilized to deduce the density scale height of
solar loops \citep{McEwan06,Duckenfield19}. Similarly,  the
multi-periodic QPPs observed on other Sun-like stars could be used
to diagnose the longitudinally-structured stellar loops
\citep{Srivastava13,Zimovets21}. It should be noted that the double
periods studied here have broad peaks, which are usually attributed
to the non-stationary property of the QPP signals and could be
interpreted by gradual change of the physical or geometrical
parameters of the flare region or by a superposition of multiple
oscillatory processes \citep[see][for recent reviews of this
issue]{Nakariakov19a,Kupriyanova20}. In our study, considering a
self-oscillation model, the broad PSD peaks in the power spectra are
supposed to be caused by the variation of inflow rates, or the
anomalous resistivity in local electric currents.

It should be stated that, from the FFT spectra shown in Figures
\ref{psd1}, \ref{psd2}, \ref{psd3}, \ref{psd4}, and \ref{psd5}, the
flare QPPs in different wavelengths are not exactly the same, but
show some deviations from each other, in particular for the short
period (P1), which could have a deviation of roughly 0.28~minutes.
This is probably because Ly\,$\alpha$, HXR, and microwave emissions
do not originate from a completely coincident source region. For
example, the HXR source of the M7.4 flare on 25 September 2011
deviates obviously from the microwave sources, as shown in
Figure~\ref{images}d, the corresponding QPPs in these two
wavelengths also shows a significant deviation. However, it is still
impossible to determine the source region of Ly\,$\alpha$ emission
due to the lack of the corresponding imaging observations. Some
instruments that have been launched recently, such as the
\emph{Extreme Ultraviolet Imager} \citep{Rochus20} and the
\emph{Spectrometer/Telescope for Imaging X-rays} \citep{Krucker20}
onboard the \emph{Solar Orbiter},  and instruments to be launched in
the near future, such as the \emph{Ly\,$\alpha$ Solar Telescope}
\citep{LiH19, Feng19} and the\emph{Hard X-ray Imager} \citep{Su19}
onboard the \emph{Advanced Space-based Solar Observatory}
\citep{Gan19,Huang19}, will provide joint imaging observations in
Ly\,$\alpha$ and HXR channels, which will help to address these
issues. Finally, we should point out that the short period (P1)
detected in the Ly\,$\alpha$, HXR, and microwave emissions might be
driven by the different mechanisms, due to their large periodic
deviation. To answer this question, a new technique is needed to
reduce the error/uncertainty of the detected period, such as the
Solar Bayesian Analysis Toolkit \citep{Anfinogentov21}.


\begin{acks}
The authors would like to thank the anonymous reviewer for their
valuable comments. We thank the teams of GOES/EUVS, GOES/XRS,
RHESSI, \emph{Fermi}, NoRP, and NoRH for their open data use policy.
This work is supported by NSFC under grants 12073081, 11973092,
11921003, 11973012, U1731241, as well as CAS Strategic Pioneer
Program on Space Science, Grant No. XDA15052200, XDA15320103, and
XDA15320301. L. Feng also acknowledges the Youth Innovation
Promotion Association for financial support. L. Lu and D. Li are
also supported by the CAS Key Laboratory of Solar Activity
(KLSA202113, KLSA202003). The Laboratory No. 2010DP173032.
\end{acks}

\section*{Declarations}

\textbf{Conflict of Interest} The authors declare that they have no conflict of interest.

%
%
 \bibliographystyle{spr-mp-sola}
 \bibliography{myreferences}
%
%
%
%

\end{article}
\end{document}